\documentclass{article}
\usepackage{spconf,amsmath,epsfig,amsfonts,graphicx,url,times,mathrsfs,color,cite,hhline}
\usepackage[bottom]{background}
\SetBgContents{Published in \emph{Proceedings of the 2024 IEEE International Conference on Multimedia and Expo, Niagara Falls, Ontario, July 15-19, 2024.}}
\SetBgVshift{1.1cm}
\SetBgOpacity{1.0}
\SetBgColor{black}
\SetBgScale{1.0}

\newcommand{\myto}{\mathrm{~to~}}

\let\OLDthebibliography\thebibliography
\renewcommand\thebibliography[1]{
  \OLDthebibliography{#1}
  \setlength{\parskip}{0pt}
  \setlength{\itemsep}{0pt plus 0.3ex}
}

\begin{document}\sloppy

\title{Why some audio signal short-time Fourier transform coefficients have nonuniform phase distributions}
\name{Stephen D. Voran}
\address{Institute for Telecommunication Sciences, Boulder, Colorado, USA, svoran@ntia.gov}

\ninept

\maketitle

\begin{abstract}
The short-time Fourier transform (STFT) represents a window of audio samples as a set of complex coefficients.
These are advantageously viewed as magnitudes and phases and the overall distribution of phases is very often assumed to be uniform.
We show that when audio signal STFT phase distributions are analyzed per-frequency or per-magnitude range, they can be far from uniform.
That is, the uniform phase distribution assumption obscures significant important details.
We explain the significance of the nonuniform phase distributions and how they might be exploited, derive their source, and explain why the choice of the STFT window shape influences the nonuniformity of the resulting phase distributions.
\end{abstract}

\begin{keywords}
DFT, noise reduction, phase distribution, phase recovery, source separation, speech enhancement, STFT 
\end{keywords}

\vspace{-2mm}
\section{Introduction}
\label{sec:intro}
\vspace{-2mm}
The discrete-time short-time Fourier transform (STFT) is a ubiquitous tool for analysis and processing of audio signals. It represents a windowed sequence of $N$ real time-domain audio samples, $x_i$, as a length $N$ sequence of complex numbers $X_k$. 
These coefficients are often decomposed into magnitude $|X_{k}|$, and phase $\phi_{k}=\angle(X_{k})$, because the squared-magnitude can be related to the power of the audio signal at a given time and frequency.

Magnitudes provide a very good baseline description of an audio signal and significant work on enhancement, separation, and classification has been accomplished using magnitudes alone.
Phase values describe how power is distributed between sinusoidal and cosinusoidal components and can be viewed as refinement information that completes the signal description and is required for exact signal reconstruction.
Research incorporating phases or complex coefficients continues to emerge and example results can be found in
\cite{Laroche1999,Parry2007,Paliwal2008,Atlas2011,GerkmannTASLP2014,Mowlaee2015,GerkmannPhaseTutorial, MOWLAEE20161,Williamson2017,Magron2018,Magron2018A,Oikawa2018,Williamson2020,Plumbley2021,Fevotte2021,Fevotte2022}.

It is commonly assumed that STFT coefficient phases $\phi_k$ are uniformly distributed. For example, this assumption is found in \cite{Parry2007,Magron2018,Mowlaee2015,MOWLAEE20161,Magron2018A,Fevotte2021,Fevotte2022}, \cite{LoizouText}, p.101 and \cite{VincentText}, p. 447. In \cite{VoranWASPAA2017} phases were measured and the \emph{global} distribution was found to be approximately uniform.
It is not hard to argue for the uniform phase distribution assumption.
Natural audio signals evolve in time and even relatively stable tonal components typically align differently with subsequent STFT windows as time passes, thus appearing more sinusoidal in some windows, more cosinusoidal in others, and in-between for yet other windows. Across any diverse collection of natural audio signals this distribution of alignments will be uniform.
But it turns out that this does not mean that all STFT coefficients will have uniform phase distributions.

Instead, our analysis of \emph{individual} STFT coefficients and our analysis by coefficient \emph{magnitude} both reveal highly nonuniform phase distributions.
In the next section we demonstrate these nonuniform phase distributions across a broad range of audio signals and we note trends with respect to window shape. In Section \ref{sec:practical} we demonstrate the significance of the nonuniform distributions and in Section \ref{sec:der} we derive their source.
Then in Section \ref{sec:tone2audio} we connect all these results back to the original audio demonstrations of Section \ref{sec:demo}. 

Significant work has been done to develop and analyze \emph{conditional} distributions for STFT phase \cite{Laroche1999,GerkmannTASLP2014,Mowlaee2015,Magron2018}.
These distributions model phases at a given time and frequency relative to an adjacent time or frequency or both, thus modeling the \emph{evolution} of phase across time or frequency.
The distributions in this paper are much more fundamental --- they are the \emph{absolute} (not conditional) phase distributions for each STFT coefficient.
Note also that the nonuniform distributions shown, analyzed, and explained in this paper are not STFT artifacts that would be removed or explained by the application of the baseboard transform as described in \cite{GerkmannTASLP2014}. Instead, the distributions are fundamental STFT properties that are not directly connected to the STFT stride.

\vspace{-2mm}
\section{Audio signals can produce nonuniform STFT coefficient phase distributions}
\label{sec:demo}
\vspace{-1.2mm}
In this section we present STFT coefficient phase histograms for audio signals and note where these histograms are far from uniform.
We use audio from five classes: speech recorded in our lab, field recordings of live music, a compilation of sound effects (31 sounds in uncompressed files from 31 different freesound.org contributors), music from professionally-produced CDs, and MP3 encoded versions of professionally-produced CDs.
This collection represents at least 60 different recording environments and equipment chains.
We use over five minutes of audio from each class.
We calculated the length $N=512$ sample STFT using the periodic Hamming window with 50\% overlap.
Window duration was 10.7 ms for speech and live music where $f_s=48$k, and 11.6 ms duration for the other three classes where $f_s=44.1$k.

\begin{figure}[!t] 
    \centering
    \includegraphics[width=85mm]{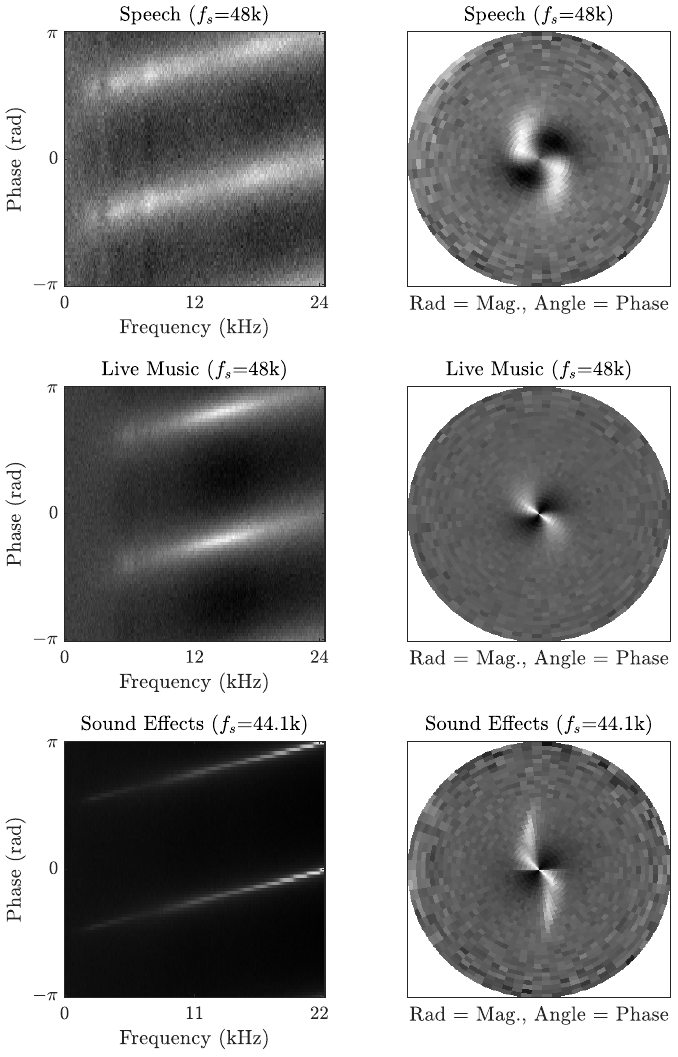}
  \caption{Audio STFT phase histogram images. Left: per-frequency histograms where each vertical stripe shows a phase histogram for a given frequency.  Right: per-magnitude histograms where each annular region shows a phase histogram for a magnitude range (smallest magnitudes at origin). White indicates highest probability.\label{fig:audioImages}}
\end{figure}

Figure \ref{fig:audioImages} shows six phase histograms as images.  The left panels show histograms by frequency.  A vertical slice at any frequency gives the phase histogram at that frequency, with white indicating the most common values. Bright white regions on dark backgrounds are prominent at many of the higher frequencies, showing that phase distributions at these frequencies are far from uniform.  The right panels show histograms for discrete magnitude ranges --- each phase histogram is an annulus. At the lowest magnitudes (near the origin) rotating from 0 to $2\pi$ takes us from black to white, then to black and to white again, clearly showing nonuniform phase distributions.
These histograms cover just three of the five audio classes, but we observed the same general trends in all five classes.
In addition, switching from Hamming to Hann windows increases uniformity while switching from Hamming to rectangular windows decreases uniformity. We are able to explain the reason for these trends in \ref{ssec:nonRect}.

We have shown that for many higher frequencies and all lower magnitudes, STFT coefficient phases are not uniformly distributed. This result is consistent across 5 audio signal classes and over 60 different audio capture and recording chains.  While these chains can alter phases of signal components and thus alter STFT phases, they cannot cause the phase distributions shown in Fig. \ref{fig:audioImages}. 

\section{these nonuniform phase distributions have practical significance}
\label{sec:practical}
\vspace{-1.5mm}
The nonuniform phase distributions observed in \ref{sec:demo} are perceptually significant even though they are often associated with lower magnitude coefficients. One demonstration of this fact follows. We selected a single STFT coefficient with nonuniform phase and added uniformly distributed phase noise from $[-\pi/4,\pi/4]$ to it and then reconstructed the original signal.  The difference (noise vs. no-noise) was clearly audible and it was easy to reliably identify the two conditions in informal blind listening. This means that it is necessary to preserve the phase values from these nonuniform distributions.

The nonuniform phase distributions observed in \ref{sec:demo} are also mathematically significant.
The distribution of a signal parameter informs the optimal strategy for quantizing that parameter. Where a phase distribution is uniform, the uniform rounding quantizer (URQ) is optimal. Where phase is not uniform, probability density function (PDF) optimized quantization will give lower RMS quantization error than the URQ \cite{Jayant}. We experimented with quantizing the STFT coefficient phase values of the sound effects signals presented in \ref{sec:demo}. We applied a single URQ across the
upper half of the band (11.025 to 22.05 kHz) and also applied a set of four PDF-optimized quantizers, each optimized for one-fourth of that band. We did this for quantizers with $2,3, \ldots, 8$ quantization cells and found that the set of four PDF-optimized quantizers reduced RMS quantization error by an average of $12\%$ compared to the single URQ.
This means that the phase values from these nonuniform distributions can be more efficiently preserved if we recognize that the distributions are nonuniform, rather than assuming them to be uniform.
More broadly, using the correct prior phase distributions could benefit algorithms that separate signals and restore proper phase values.

\vspace{-1mm}
\section{the cause of these nonuniform phase distributions}
\label{sec:der}
\vspace{-1mm}
We have shown that audio signal STFT coefficient phase distributions are often nonuniform.
This contrasts with and significantly refines the widely-held global uniform phase assumption,
so we seek to explain \emph{why} these distributions are nonuniform.
Audio signals generally have strong tonal components so we now turn to tones.

\subsection{How tone phase determines STFT phase}
\label{ssec:tonePhase}
We first derive the relationship between the phase of a tone at an arbitrary frequency $\omega_t$ and the resulting phase of an arbitrary STFT coefficient at frequency $\omega_k$.
We show that this relationship is linear when the tone is near the coefficient ($\omega_t \approx \omega_k$) but it can become increasingly nonlinear as the tone frequency and the coefficient frequency become increasingly distant.
These nonlinear relationships map \emph{uniform} distributions of tone phases to \emph{nonuniform} distributions of coefficient phases.
Thus audio signals that are composed of tones with uniformly distributed phases can produce STFT coefficients with nonuniformly distributed phases.

Using the STFT frequency sample index $k\!=\!0$ to $N\!-\!1$, window (or frame) index $t = 0, 1, ...$, and stride $N_s$, the STFT can be written
\vspace{-2mm}
\begin{align}
\label{eqn:STFT}
X_{k,t} = \sum_{i=0}^{N-1} w_i x_{(t\cdot N_s+i)} e^{-j2\pi k i/N}~,
\end{align}
where $w_i$, $i = 0$ to $N-1$ are the window values.
The STFT kernel (or bin) frequencies are located at $\omega_k=2 \pi k/N$~radians (or $f_k = kf_s/N$ Hz for sample rate $f_s$ smp/s). We use a common choice for stride, $N_s = N/2$ throughout this work. For clarity, we omit the window index $t$ going forward. Consistent with common practice, we assume $N$ is even. The audio signal $x_i$ is real, so $X_0$ (DC) and $X_{N/2}$ (Nyquist) are real and coefficients show complex conjugate symmetry about Nyquist. This means that only $X_1$ through $X_{N/2-1}$ carry unique phase information. So throughout this paper we consider only $k=1$ to $N/2-1$ without loss of generality.

We consider windows of the form
\begin{align}
\label{eqn:window}
w_i(\alpha,N_w) = (1-\alpha) - \alpha \cos{(2\pi i/N_w)}, ~~i = 0 \myto N-1.
\end{align}
Selecting window parameter $\alpha =0$ gives the rectangular window with a narrow main lobe but minimal sidelobe suppression.  Increasing $\alpha$ generally broadens the main lobe but improves sidelobe suppression and $\alpha=0.46$ and $0.50$ correspond to the Hamming and Hann windows respectively; both of these are used in the seminal phase reconstruction work \cite{GriffinLim1984}. $N_w = N-1$ produces symmetric windows while $N_w = N$ gives periodic windows which can have the desirable constant-amplitude overlap-and-add (OLA) property \cite{Rabiner&Schafer_speech}, and their square roots have the constant-power OLA property. Thus periodic Hamming and Hann windows and their square roots are common choices for audio signals. The sensitivities of STFT phase, phase reconstruction, and intelligibility to window parameters have been studied in \cite{windowsPhase} and \cite{windowsPhaseReconstruction}.

A tone with arbitrary phase \mbox{$\theta \in [-\pi,\pi)$} and frequency \mbox{$\omega_t \in [0,\pi]$} between DC and Nyquist is given by
\begin{align}
\label{eqn:tone}
x_i(\omega_t,\theta) = \cos{(\omega_t i + \theta)}, ~~i = 0 \myto N-1.
\end{align}
We consider the windowed STFT of this tone.
We wish to find the phase $\phi_k$ of each STFT coefficient $X_k$.
We insert the tone definition (\ref{eqn:tone}) into the STFT definition (\ref{eqn:STFT}), and apply trigonometric product-to-sum identities to get
\begin{align}
\label{eqn:DFT1}
X_k &=  \sum_{i=0}^{N-1} w_i \cos{(\omega_t i + \theta)} e^{-j \omega_k i}  \nonumber \\
&=\frac{1}{2}  \sum_{i=0}^{N-1} w_i \left[ e^{j((\omega_t - \omega_k) i + \theta )} + e^{-j((\omega_t + \omega_k) i + \theta)} \right]. 
\end{align}
Inserting the window definition (\ref{eqn:window}) into (\ref{eqn:DFT1}) and applying trigonometric product-to-sum identities yields twelve trigonometric sums. The sums and their solutions take the form
\vspace{-1mm}
\begin{align}
\label{eqn:trigsums}
f(\gamma,\theta,N) = \sum_{i=0}^{N-1}&\sin(\gamma i + \theta )  = 
    {s(\gamma,N)\sin((N\!\!-\!1){\gamma \over 2}\!+\theta)}, \nonumber \\
g(\gamma,\theta,N) = \sum_{i=0}^{N-1}&\cos(\gamma i + \theta )  = 
    {s(\gamma,N)\cos((N\!\!-\!1){\gamma \over 2}\!+\theta)}, \nonumber \\
\text{where } s(\gamma,N) =& 
\begin{cases}
{ {\sin(N{\gamma \over 2})} \over {\sin({\gamma \over 2})}}, -\pi \!\leq \! \gamma \! < \! 0 ~~\mathrm{or}~~0 \! < \! \gamma \! < \! 2\pi, \\
N, ~~~\gamma = 0.
\end{cases}
\end{align}
With these definitions in place, the real and imaginary parts of the STFT coefficient $X_k$ in (\ref{eqn:DFT1}) are given by
\vspace{-2mm}
\begin{align}
\label{eqn:realImag}
\Re(X_k) =  {{(1-\alpha)} \over 2} \Big[ g(\Delta^+_k,\theta,N) +  g(\Delta^-_k,\theta,N)\Big] \nonumber \\
 - {\frac{\alpha}{4}}
 \Big[  g(\Delta^+_k-\beta,\theta,N) + g(\Delta^+_k+\beta,\theta,N) \nonumber \\
 + g(\Delta^-_k-\beta,\theta,N) + g(\Delta^-_k+\beta,\theta,N) \Big], \nonumber \\
 \Im(X_k) =  {{(1-\alpha)} \over 2} \Big[ -f(\Delta^+_k,\theta,N) +  f(\Delta^-_k,\theta,N)\Big] \nonumber \\
 - {\alpha \over 4}
 \Big[  -f(\Delta^+_k-\beta,\theta,N) - f(\Delta^+_k+\beta,\theta,N) \nonumber \\
 + f(\Delta^-_k-\beta,\theta,N) + f(\Delta^-_k+\beta,\theta,N) \Big], \nonumber \\
\mathrm{where~~} \Delta^\pm_k(\omega_t) =  \omega_t \pm \omega_k \mathrm{~~and~~}
\beta =  2 \pi/N_w.
\end{align}
Finally, the phase of the coefficient is found via the four-quadrant arctangent function:
\begin{align}
\label{eqn:atan}
\phi_k = \angle(X_k) =  \arctan(\Re(X_k),\Im(X_k)),
\end{align}
(where $\arctan(0,0) := 0$). Equations (\ref{eqn:realImag}) and (\ref{eqn:atan}) provide closed-form expressions for the STFT phase $\phi_k$ produced by a windowed tone with arbitrary frequency and phase.
Noting that $s(2\omega_k,N) = 0$ and $s(0,N) = N$, it is not difficult to show that when the tone frequency matches an STFT coefficient frequency, the coefficient phase and the tone phase will match (rectangular window or periodic windows) or nearly match (symmetric windows):
\begin{align}
\label{eqn:toneOnBin}
\omega_t = \omega_k \implies 
\begin{cases}
\phi_k = \theta, ~\mathrm{when}~ \alpha\! = \!0, \\
\phi_k = \theta, ~\mathrm{when}~ \alpha\! \neq\! 0 ~\mathrm{and}~N_w\! =\! N, \\
\phi_k \approx \theta, ~\mathrm{when}~ \alpha \!\neq\! 0 ~\mathrm{and}~N_w\! \!= N\!-\!1. \\
\end{cases}
\end{align}

\subsection{Rectangular window case}
\label{ss:rectangularWindow}
To gain initial insight we now consider the case of the rectangular window ($\alpha = 0$).
We can best understand the general relationship between $\theta$ and $\phi_k$ by rewriting (\ref{eqn:realImag}) to isolate the dependence on $\theta$:
\begin{align}
\label{eqn:realImagRect}
\Re(X_k) =  a_\Re(\omega_t,&\ \omega_k,N) \cos(\theta) + b_\Re(\omega_t,\omega_k,N) \sin(\theta), \nonumber \\
\Im(X_k) =  a_\Im(\omega_t,&\ \omega_k,N) \cos(\theta) + b_\Im(\omega_t,\omega_k,N) \sin(\theta), \nonumber \\
a_\Re(\omega_t,\omega_k,N) &=\ \scriptstyle\frac{1}{2}\textstyle
\left(+g(\Delta_k^+,0,N) + g(\Delta_k^-,0,N)\right), \nonumber \\
b_\Re(\omega_t,\omega_k,N) &=\ \scriptstyle\frac{1}{2}\textstyle
\left(-f(\Delta_k^+,0,N) - f(\Delta_k^-,0,N)\right), \nonumber \\
a_\Im(\omega_t,\omega_k,N) &=\ \scriptstyle\frac{1}{2}\textstyle
\left(-f(\Delta_k^+,0,N) + f(\Delta_k^-,0,N)\right), \nonumber \\
b_\Im(\omega_t,\omega_k,N) &=\ \scriptstyle\frac{1}{2}\textstyle
\left(-g(\Delta_k^+,0,N) + g(\Delta_k^-,0,N)\right).
\end{align}
Using the relation for linear combinations of cosines and sines, we can rewrite (\ref{eqn:realImagRect}) as
\begin{align}
\label{eqn:realImagRect2}
\Re(X_k)  = c_\Re(\omega_t,\omega_k,N) \cos(\theta + \zeta_\Re(\omega_t,\omega_k,N)), \nonumber \\
\Im(X_k)  = c_\Im(\omega_t,\omega_k,N) \cos(\theta + \zeta_\Im(\omega_t,\omega_k,N)), \nonumber \\
\mathrm{where}~ c_\Re = \sqrt{a_\Re^2 + b_\Re^2}, ~~\zeta_\Re = \arctan(a_\Re,-b_\Re), \nonumber \\
c_\Im = \sqrt{a_\Im^2 + b_\Im^2}, ~~\zeta_\Im = \arctan(a_\Im,-b_\Im). 
\end{align}
Finally, using (\ref{eqn:realImagRect2}) in (\ref{eqn:atan}) shows that when $\omega_t \neq \omega_k$, STFT phase is
\begin{align}
\label{eqn:atanRect}
\phi_k =   \arctan(\,(c_\Re/c_\Im)\cos(\theta+\zeta_\Re)\,,\,\cos(\theta+\zeta_\Im)\,).
\end{align}

\subsection{Linear and nonlinear phase relationships}
\label{ss:linAndNon}
Equations (\ref{eqn:toneOnBin}) and (\ref{eqn:atanRect}) give the relationship between the phase $\theta$ of a tone with arbitrary frequency $\omega_t$  and the resulting phase $\phi_k$ of an  arbitrary STFT coefficient at frequency $\omega_k$ in the case of the rectangular window.
These results make it clear that STFT phase is periodic in tone phase, as expected.
And they show that a wide variety of relationships between tone phase and STFT phase are possible.
These relationships are driven by the angles $\zeta_\Re$ and $\zeta_\Im$ as well as the ratio $c_\Re/c_\Im$.  These are in turn are driven by $\omega_t$, $\omega_k$, and $N$.

\begin{figure}[h] 
    \centering
    \includegraphics[width=85mm]{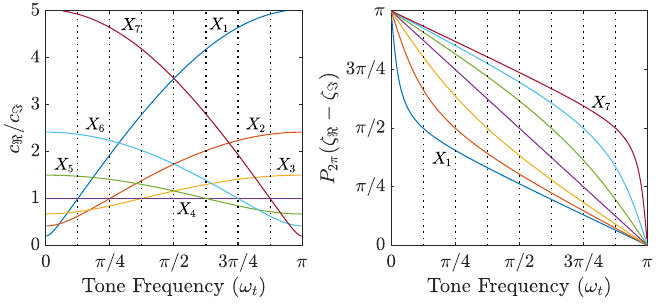}
  \caption{Example shows how tone frequency $\omega_t$ drives $c_\Re/c_\Im$ and $P_{2\pi}(\zeta_\Re\!-\!\zeta_\Im)$ per (\ref{eqn:atanRect}) for each of the seven STFT coefficients of interest when $N=16$. STFT coefficient frequencies shown by dotted vertical lines. Right panel uses same color coding as left panel. \label{fig:cAndZVSwt}}
\end{figure}

Figure \ref{fig:cAndZVSwt} shows these relationships for the case $N=16$.
The left panel of Fig. \ref{fig:cAndZVSwt} uses seven colors (as annotated) to show $c_\Re/c_\Im$ for all seven STFT bins of interest ($X_k$, $k = 1$ to $7$) as the tone frequency $\omega_t$ moves from $0$ to $\pi$. 
The right panel is organized in the same manner and shows $P_{2\pi}(\zeta_\Re\!-\!\zeta_\Im)$, where $P_{2\pi}(\cdot)$ maps angles into a principal angle in $[0,2\pi)$. This is an efficient way to visually appreciate a key influence of $\zeta_\Re$ and $\zeta_\Im$ in (\ref{eqn:atanRect}).   

In both panels each of the seven frequencies $\omega_k$ are marked with a dotted vertical line. This makes it easy to see that when $w_t = w_k$,  the curves for $X_k$ give $c_\Re/c_\Im = 1$ and  $P_{2\pi}(\zeta_\Re\!-\!\zeta_\Im) = \pi/2$, so (\ref{eqn:atanRect}) produces $\phi_k = \theta$.
More generally, the smooth nature of Fig. \ref{fig:cAndZVSwt} dictates that the relationship between $\theta$ and $\phi_k$ is approximately linear when $w_t \approx w_k$.
But if $w_t$ moves away from $w_k$, then $c_\Re/c_\Im$ moves away from $1$, $P_{2\pi}(\zeta_\Re\!-\!\zeta_\Im)$ moves away from $\pi/2$ and given the roles these variables play in (\ref{eqn:atanRect}), it is clear that the relationship between $\theta$ and $\phi_k$ can become highly nonlinear.
Fig. \ref{fig:cAndZVSwt} illuminates the case $N\!\!=\!\!16$, but the trends shown hold for general values of $N$.

\begin{figure}[htp] 
    \centering
    \includegraphics[width=85mm]{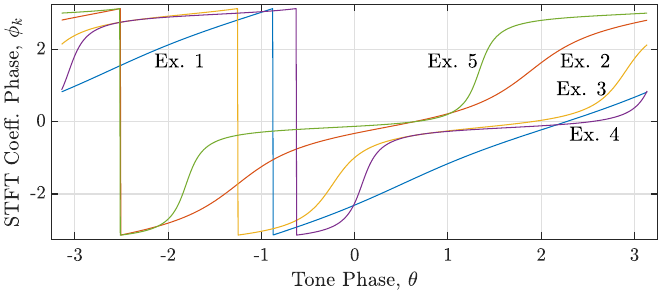}
  \caption{Five example relationships between tone phase $\theta$ and STFT coefficient phase $\phi_k$. Conditions for each example given in Table \ref{table:examples}. \label{fig:angeRlnExamples}}
\end{figure}

Table \ref{table:examples} and Fig. \ref{fig:angeRlnExamples} show five example relationships between tone phase $\theta$ and the phase of an STFT coefficient $\phi_k$. In Example 1 the tone frequency $\omega_t$ is very close to the coefficient frequency $\omega_k$ and the resulting phase relationship is nearly linear.  In examples 2, 3, and 4 the tone frequency moves away from the coefficient frequency and the phase relationship becomes increasingly nonlinear.  Example 5 shows that changing $N$ from 512 to 2048 shifts the phase relationship but does not alter its nonlinear nature. 

\vspace{-5mm}
\begin{table}[h] 
\caption{Conditions for five examples shown in Fig. \ref{fig:angeRlnExamples}. Tone frequency $\omega_t$, STFT coefficient frequency $\omega_k$, and their difference $\Delta_k^- = \omega_t-\omega_k$ all in radians, STFT length $N$ in samples, $c_\Re$, $c_\Im$, $\zeta_\Re$, and $\zeta_\Im$ given in (\ref{eqn:realImagRect2}) and discussed in \ref{ss:linAndNon}.}
\centering
\begin{tabular}{|l|l|l|l|l|l|l|}
\hline
Ex. & $~~\omega_k$    & ~~$\omega_t$ &~~~$\Delta_k^-$   &\!\! $N$ \!\!&  \!\!$c_\Re/c_\Im$ \!\!& \!\!$P_{2\pi}(\zeta_\Re \!\! - \!\! \zeta_\Im)\!\!$  \\ \hline
1  & $0.90\pi$  & $0.88\pi$ & \!$-0.02\pi$  & 512  & 1.18    & $0.52\pi$        \\ \hline
2  & $0.90\pi$  & $0.80\pi$ & \!$-0.10\pi$  & 512  &  1.93   &  $0.57\pi$       \\ \hline
3  & $0.90\pi$  & $0.75\pi$ & \!$-0.25\pi$  & 512  & 3.25   &  $0.66\pi$        \\ \hline
4  & $0.90\pi$ & $0.45\pi$ & \!$-0.45\pi$  & 512  & 4.73    &   $0.79\pi$       \\ \hline
5 & $0.90\pi$  & $0.45\pi$ & \!$-0.45\pi$  & \! 2048 \! & 4.82     &   $0.77\pi$     \\ \hline
\end{tabular}
\label{table:examples}
\vspace{-4mm}
\end{table}

\vspace{-2mm}
The nonlinear mappings of tone phase $\theta$ to STFT coefficient phase $\phi_k$
developed here and exemplified in Fig. \ref{fig:angeRlnExamples} allow a uniform distribution of tone phases to produce a nonuniform distribution of STFT coefficient phases, as seen in \ref{sec:demo}.

\vspace{-3mm}
\subsection{Each STFT coefficient has intrinsic phase values} 
\label{ssec:intrinsicValues}
Figure \ref{fig:audioImages} is based on significant quantity and diversity of audio. Yet the left panels show very clear specific trends of higher phase density locations (white) that increase with STFT coefficient frequency. These trends suggest that these locations of higher phase density are intrinsic to each STFT coefficient. We now show that in the case of tonal signal components and the rectangular window, each STFT coefficient does indeed have four \emph{intrinsic} locations for phase peaks. This explains why tonal components of many different frequencies all contribute to a single set of phase peaks.

When the uniformly distributed random variable tone phase $\theta$ is transformed to produce the  STFT coefficient phase $\phi_k$ (see \ref{ss:linAndNon}), we can find the probability density function (pdf) of this new random variable as follows\cite{Papoulis}.
We use $\phi_k = F(\theta)$ to represent (\ref{eqn:trigsums}), (\ref{eqn:realImag}), and (\ref{eqn:atan}) for $\alpha=0$ and fixed values of $N$, $k$, and $\omega_t$. We use $G$ as shorthand for the inverse function \mbox{$\theta = G(\phi_k) = F^{-1}(\phi_k)$}. Then the pdf for $\phi_k$ is given by 

\begin{align}
\label{eqn:phipdf}
&f_{\phi_k}(\phi_k) = f_\theta(G(\phi_k))\left| {F^\prime(G(\phi_k))}\right|^{-1}= \\ \nonumber
&\left| \frac
{s^2(\Delta^+_k)\!+\!s(\Delta^+_k)s(\Delta_k^-)2\cos \left( (N\!\!-\!\!1)\omega_t\!+\!2G(\phi_k) \right) \!+\!s^2(\Delta^-_k)} 
{2\pi\left(-s^2(\Delta^+_k)\!+\!s^2(\Delta^-_k)\right)}
\right|,
\end{align}
where $s(\gamma)$ is abbreviated notation for the function $s(\gamma,N)$ defined in (\ref{eqn:trigsums}). To find the peaks in this pdf we first note that the only $\phi_k$ dependence is in the center term of the numerator.  Next we consider the signs of the first two factors of this center term.  It is not difficult to show that
\begin{align}
\label{eqn:casesAndSigns}
s(\Delta^+_k\!,N)s(\Delta^-_k,N) &\leq 0, ~\mathrm{when}~ \omega_t < \omega_k,  \nonumber \\
s(\Delta^+_k\!,N)s(\Delta^-_k,N) &\geq 0, ~\mathrm{when}~ \omega_t > \omega_k.  
\end{align}
It follows that the pdf given in (\ref{eqn:phipdf}) is maximized w.r.t $\phi_k$ by minimizing $\cos \left( (N\!\!-\!\!1)\omega_t\!+\!2G(\phi_k) \right)$ when  $\omega_t < \omega_k$, and by maximizing $\cos \left( (N\!\!-\!\!1)\omega_t\!+\!2G(\phi_k) \right)$ when  $\omega_t > \omega_k$.
In other words, (\ref{eqn:phipdf}) is maximized when $\phi_k = \phi_k^*$ such that
\begin{align}
\label{eqn:thetaStar}
G(\phi_k^*) &= P_\pi\!\left(\pi-(N-1)\frac{\omega_t}{2}  \pm \frac{\pi}{2}\right), &\mathrm{when}~ \omega_t < \omega_k, \nonumber \\
G(\phi_k^*) &= P_\pi\!\left(\frac{\pi}{2}-(N-1)\frac{\omega_t}{2}  \pm \frac{\pi}{2}\right), &\mathrm{when}~ \omega_t > \omega_k,
\end{align}
where $P_\pi(\cdot)$ maps angles into a principal angle in $[-\pi,\pi)$.
We apply $F$ to both sides of (\ref{eqn:thetaStar}) to convert tone phase to STFT phase:
\begin{align}
\label{eqn:phiStar1}
\phi_k^* &=   P_\pi\!\left( \frac{\omega_k \pm \pi}{2}\right), &\mathrm{when}~ \omega_t < \omega_k, \\
\label{eqn:phiStar2}
\phi_k^* &=   P_\pi\!\left( \frac{\omega_k +\pi \pm \pi}{2}\right), &\mathrm{when}~ \omega_t > \omega_k.
\end{align}
These are the locations of the peaks in the pdf of the STFT phase $f_{\phi_k}(\phi_k)$. These locations do \emph{not} depend on the exact value of $\omega_t$, they only depend on whether $\omega_t$ is above or below $\omega_k$.
This means that tones of all frequencies below $\omega_k$ contribute to the same two peaks in the phase distribution
and all tones above $\omega_k$ contribute to the other two peaks.

\begin{figure}[!ht] 
    \centering
    \includegraphics[width=85mm]{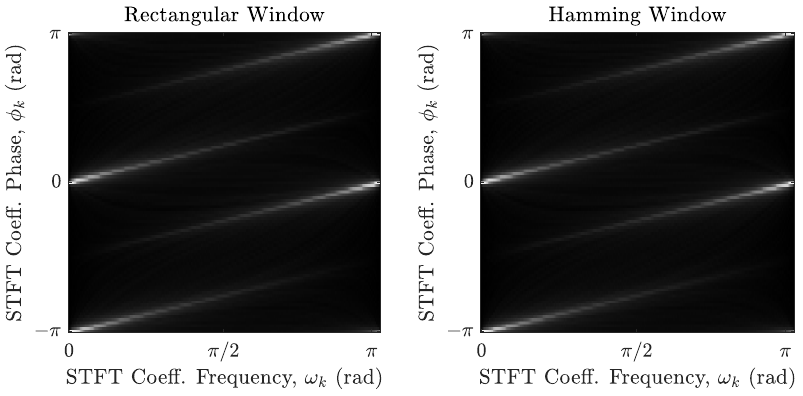}
    \vspace{-3mm}
  \caption{STFT phase histogram image shows that tones with uniformly distributed frequencies and phases produce nonuniformly distributed STFT phases. White is highest probability.\label{fig:toneImage}}
\end{figure}

Figure \ref{fig:toneImage} provides empirical confirmation. We created $10^4$ tones by drawing a uniformly distributed random frequency \mbox{$\omega_t \in [0,\pi]$} and  a uniformly distributed random phase 
\mbox{$\theta \in [-\pi,\pi)$} for each tone.
We then applied a length $N\!\!=\!512$ rectangular-windowed STFT to each tone. The left side of Fig. \ref{fig:toneImage} shows the resulting phase histogram calculated across all $10^4$ results. The right side shows the same experiment done with the periodic Hamming window and is visually identical (and numerically very close) to the left side.
The phase peaks given by (\ref{eqn:phiStar2}) show most prominently in Fig. \ref{fig:toneImage} (as white lines) near the left edges of the images since the majority of the tones have frequencies above these STFT bins.
And the phase peaks given by (\ref{eqn:phiStar1}) are most prominent  near the right edges since the majority of the tones have frequencies below these STFT bins.
Both the rectangular and periodic Hamming window experiments shown in Fig. \ref{fig:toneImage} agree with the rectangular window derivation results in \ref{ssec:intrinsicValues}.
In addition, we have empirically found that the phase peaks follow (\ref{eqn:phiStar1}) and (\ref{eqn:phiStar2}) for $0 \leq \alpha \leq 0.49$ but not for $\alpha=0.5$ (the periodic Hann window). We repeated the experiment with noise in place of tones and obtained uniform phase distributions, as expected.

\subsection{Influence of window shape} 
\label{ssec:nonRect}
Window shapes influence STFT phase phase values through their sidelobe suppression properties.
Consider a signal that has 
some tonal spectral components near $\omega_k$ and some that are far from $\omega_k$ and all of these spectral components have uniformly distributed phases. From \ref{ss:linAndNon} and \ref{ssec:intrinsicValues} we know that any spectral components close to $\omega_k$ would cause $X_k$ to have a uniform phase distribution, while spectral components away from $\omega_k$ would lead to peaks at a pair of phase values.
So the resulting phase distribution for $X_k$ will be determined by the relative powers of the near and far spectral components \emph{and} by the sidelobe suppression of the window.
A window with more suppression will better block the far components and will result in a more uniform phase distribution.  A window with less suppression will allow the far components to more strongly influence $X_k$, resulting in a less uniform phase distribution. 

\begin{figure}[t]
    \centering
    \includegraphics[width=85mm]{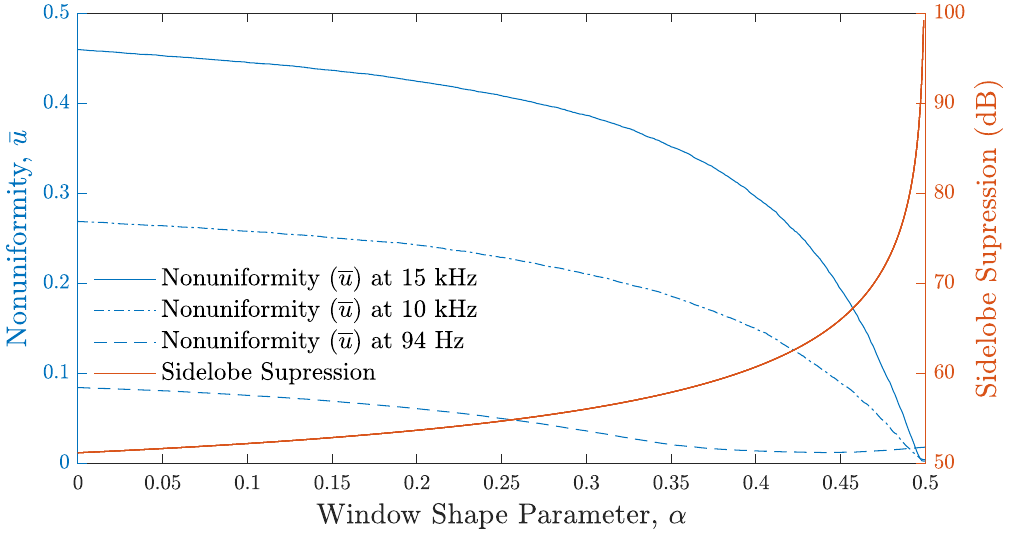}
  \caption{Increasing window shape parameter (see (\ref{eqn:window}) and \ref{ssec:tonePhase}) increases sidelobe suppression and thus decreases nonuniformity $\overline{u}$ of phase distributions.\label{fig:vsAlpha}}
\end{figure}

Figure \ref{fig:vsAlpha} gives an empirical example for the live music audio class used in \ref{sec:demo}. It shows a measure of phase distribution nonuniformity
$\overline{u}$ for three different STFT coefficients and an average measure of sidelobe suppression at $\pi/2$ radians frequency offset.
As $\alpha$ increases from $0$ (rectangular window) through $0.46$ (periodic Hamming) and on to $0.5$ (periodic Hann), sidelobe suppression increases and nonuniformity decreases, as expected, and consistent with our observations on rectangular, Hamming, and Hann windows in \ref{sec:demo}.

The quantitative measure of nonuniformity $\overline{u}$ used in Fig. \ref{fig:vsAlpha} is the normalized earth mover's distance (EMD) \cite{EMD} between the measured phase distribution $\{p_m\}_{m=0}^{M-1}$ and the uniform distribution,
\vspace{-2mm}
\begin{align}
\label{eqn:ubar}
\overline{u}\left(\{p_m\}\strut_{m=0}^{M-1}\right) = \frac{1}{u_0}\sum_{k=0}^{M-1}\left| \sum_{m=0}^k\left(p_m - \frac{1}{M}\right)\right|.
\end{align}
The normalization factor $u_0$ is the EMD between the uniform distribution and the deterministic case where all probability mass is in the central bin.
Thus the range of $\overline{u}$ is from  0.0 (when $p_m$ is uniform) to 1.0 (when $p_m$ is deterministic).

Another perspective on how window shapes influence STFT phase values is seen by considering the convolution theorem that equates time-domain multiplication of signal and window with frequency-domain circular convolution of signal STFT and window STFT: STFT$(x \cdot w) = X * W$. This formulation emphasizes that the change imparted by switching from a rectangular window (where $W$ is the convolutional identity) to a more general window is circular convolution with the STFT of that new window ($W$). The STFT of a periodic window is real and the STFT of a symmetric window has only a small imaginary part which scales down with $1/N$ as window length grows. So $X * W \approx X * \Re(W)$. That is, these windows modify the real and imaginary parts of $X$ in similar fashion.

\section{Tone results explain general audio results}
\label{sec:tone2audio}

In \ref{ss:linAndNon} and \ref{ssec:intrinsicValues} we derived the mechanism by which tones with uniformly distributed phases can produce nonuniform STFT phase distributions with peaks at specific phases. In \ref{ssec:nonRect} we showed how the sidelobe suppression of windows influence phase distributions. Together, these results explain how more general audio signals produce nonuniform phase distributions for some STFT coefficients, as clearly shown in Fig. \ref{fig:audioImages}.

Audio signals can be modeled as mixtures of tonal and noise-like components, with tonal components typically having far greater power than the noise-like components.
So audio signals produce results similar to the tone results, but with variations due to the audio signals' specific distributions of tonal and noise-like components across frequency and time.
Spectra of audio signals typically show large dynamic range with spectral peaks many tens of decibels above spectral valleys.
Signals often have much less power at higher frequencies than at lower frequencies and if proper anti-aliasing filters are used, then the power near Nyquist frequency will be very low. This large dynamic range means that tonal components at lower frequencies can dominate the response of higher frequency STFT coefficients, even when windows with good sidelobe suppression are used.

This explains the common trend across the signals described in \ref{sec:demo} and shown in Fig. \ref{fig:audioImages}--- nonuniformity (manifested as white bands) is more common at the higher STFT frequencies ($\omega_k$) where the power of the lower frequency tonal components ($\omega_t<\omega_k$) may dominate tonal components near $\omega_k$ in spite of significant attenuation provided by the window sidelobe suppression. As shown in \ref{ss:linAndNon}, there are nonlinear mappings between the phases of the lower-frequency tonal components and resulting phases of the distant higher-frequency STFT coefficients. Further, these nonlinear mappings cause many different frequencies to all contribute to a single pair of phase peaks (see \ref{ssec:intrinsicValues}) located at $P_\pi(\pi k/N \pm \pi/2)$. These derived peak locations agree with the empirically found areas of highest probability shown in white in Fig. \ref{fig:audioImages} (audio with periodic Hamming window) and in Fig. \ref{fig:toneImage} (tones with rectangular and periodic Hamming windows).

Since the peak locations are different for each STFT coefficient, it is possible for the distribution over \emph{all} coefficients to be roughly uniform.  But even in cases where the uniform global phase assumption is approximately correct, that assumption is far from ideal because taking such a high-level view obscures the explainable and exploitable highly nonuniform distributions that are easily seen in a per-coefficient (or per magnitude level) analysis.

\section{Summary}
\label{sec:disc}  
We have demonstrated that audio signals can produce STFT coefficients with nonuniform phase distributions.
This is a significant refinement over the common assumption of a global uniform distribution.
We described the perceptual and mathematical significance of the nonuniform phase distributions.
We derived the mechanism by which the phases of tonal audio components are nonlinearly mapped to STFT phases and explained how this produces nonuniform STFT phase distributions when the audio signal spectra and the STFT window characteristics combine to allow distant tonal components to dominate a given STFT coefficient.
In summary, an audio signal will produce nonuniformly distributed STFT phases $\phi_k$ when
\begin{itemize}
     \item  There is a nonlinear relationship between tone phases and STFT coefficient phase $\phi_k$ for tonal components that are located in some frequency neighborhood $T$ \emph{and}
     \item Window sidelobe suppression and audio spectral content combine to allow tonal components in that same neighborhood $T$ to dominate the response of $X_k$ and hence $\phi_k$. 
 \end{itemize}
Finally, we suggest that when audio signal processing requires a prior phase distribution, it could be worthwhile to experiment with using a per-frequency, per-band, or per-magnitude level nonuniform prior to see if that additional specificity leads to improved performance or efficiency compared to simply adopting the uniform prior.

\bibliographystyle{IEEEbib}
\bibliography{sources}

\end{document}